\documentclass[%
 reprint, 
 amsmath,amssymb,
 aps,
prl,
floatfix,
]{revtex4-1}

\usepackage{graphicx}
\usepackage{epsfig} 
\usepackage{dcolumn}
\usepackage{bm}
\usepackage{epstopdf}
\usepackage{multirow}
\usepackage{amsmath}
\usepackage{booktabs}
\usepackage{color}
\usepackage{gensymb}
\usepackage{kantlipsum}  
\usepackage{hyperref}
\usepackage{braket}
\usepackage{physics}
\usepackage{url}
\usepackage[utf8]{inputenc}
\usepackage[T1]{fontenc}
\usepackage{mathptmx}
\usepackage{lineno}

\begin{document}

\preprint{APS/123-QED}

\title{Jet-Loaded Cold Atomic Beam Source for Strontium} 



\author{Minho Kwon}
\thanks{These authors contributed equally.}
\affiliation{Department of Physics, Columbia University, 538 West 120th Street, New York, New York 10027, USA}
\author{Aaron Holman}
\thanks{These authors contributed equally.}
\affiliation{Department of Physics, Columbia University, 538 West 120th Street, New York, New York 10027, USA}
\author{Quan Gan}
\affiliation{Department of Physics, Columbia University, 538 West 120th Street, New York, New York 10027, USA}
\author{Chun-Wei Liu}
\affiliation{Department of Physics, Columbia University, 538 West 120th Street, New York, New York 10027, USA}
\author{Matthew Molinelli}
\affiliation{Department of Physics, Columbia University, 538 West 120th Street, New York, New York 10027, USA}
\author{Ian Stevenson}
\affiliation{Department of Physics, Columbia University, 538 West 120th Street, New York, New York 10027, USA}
\author{Sebastian Will}
\thanks{The author to whom correspondence may be addressed: sebastian.will@columbia.edu}
\affiliation{Department of Physics, Columbia University, 538 West 120th Street, New York, New York 10027, USA}


\date{\today}

\begin{abstract}
We report on the design and characterization of a cold atom source for strontium (Sr) based on a two-dimensional magneto-optical trap (MOT) that is directly loaded from the atom jet of a dispenser. We characterize the  atom flux of the source by measuring the loading rate of a three-dimensional MOT. We find loading rates of up to $10^8$ atoms per second. The setup is compact, easy to construct, and has low power consumption. It addresses the long standing challenge of reducing the complexity of cold beam sources for Sr,  which is relevant for optical atomic clocks and quantum simulation and computing devices based on ultracold Sr.
\end{abstract}

\pacs{}

\maketitle 


\section{Introduction}

In ultracold quantum science the impact of atoms with two valence electrons, such as strontium (Sr) and ytterbium (Yb), has dramatically increased over the past years 
\cite{takasu2003spin, takamoto2005optical, fukuhara2007degenerate,ludlow2008srPaper, scazza2014observation, McGuyer2014,dunning2016recent,cooper2018alkaline,norcia2018microscopic, kondov2019molecular, wilson2022trapping, jenkins2022ytterbium}. With a level structure that features singlet and triplet electronic states, these atoms have a wide gamut of internal transitions \cite{loftus2001YtterbiumThesis, boyd2007strontium}, including transitions with broad linewidths (several MHz) that are well-suited for highly effective laser cooling; with narrow linewidths (tens of kHz) that are used for laser cooling to Doppler temperatures in the microkelvin range; and with ultranarrow linewidths (less than Hz) that enable highly coherent quantum operations. Combined with the presence of magic wavelengths \cite{ye2008quantum, Guo_2010, safronova2015extracting, Yamamoto_2016, ma2022universal}, tune-out wavelengths \cite{Heinz2020}, and optically trappable Rydberg states \cite{madjarov2020high,wilson2022trapping}, which all are a result of the rich two-electron level structure, Sr and Yb have emerged as important atoms for optical atomic clocks, quantum simulators, and quantum computers.

Focusing on Sr, several groundbreaking advances in the past few years have shown its exceptional potential for quantum science and technology. Today's most accurate atomic clocks are using fermionic $^{87}$Sr trapped in optical lattices \cite{campbell2017fermi,oelker2019demonstration}. Building on these advances, the concept of Sr atomic clocks is currently combined with optical tweezer trapping technology, showing competitive clock performance \cite{madjarov2019atomic,norcia2019seconds,young2020half}. Sr atomic clocks are a potential candidate for an upcoming redefinition of the second \cite{Bloom2014, Beloy2021}, replacing the definition from 1967 based on a microwave transition in cesium. In addition, optical tweezer platforms utilizing $^{87}$Sr and $^{88}$Sr are showing great promise for quantum computing \cite{pagano2019fast, pagano2022error}, including the demonstration of highly coherent nuclear spin qubits \cite{barnes2022assembly} and Bell state generation with extremely high fidelity \cite{madjarov2020high}. 

To realize the promise of Sr platforms on a broad scale and allow for the construction of deployable Sr-based quantum devices, robust and compact hardware for the preparation of ultracold Sr is critical. In this context, Sr atomic sources face particular technical challenges. Due to its high melting (769 $^\circ$C) and boiling point (1,384 $^\circ$C), Sr tends to stick to viewports and the inner walls of room-temperature vacuum chambers. As a result, sources based on a vapor cell, which are highly functional for alkali atoms, such as Rb and Cs, cannot be realized for Sr (similar to Yb, Er, Dy). Instead, Sr sources often rely on an effusive oven combined with a Zeeman slower. Such slowers have a cold atom flux of up to $10^9$ atoms per second, but are large (typically 1 m long) and use electromagnets that are power-hungry and intricate to build \cite{boyd2007strontium,nicholson2015atomicclock,yang2015high}. It has been shown that permanent magnets can be used to replace electromagnets, while the overall dimensions of the slower remain large \cite{reinaudi2012dynamically}. A more compact solution that combines a Zeeman slower with transverse cooling is available commercially \cite{AOSense}, reaching a trappable cold atom flux of about $10^9$ atoms per second \cite{baumgrtner2017ANA}, but is technologically complex, costly, and difficult to service.

Two-dimensional (2D) magneto optical traps (MOTs), creating an atomic beam via transverse laser cooling in two directions, are a popular alternative source concept \cite{dieckmann1998two,schoser2002intense,chaudhuri2006realization,catani2006intense}. For alkali atoms, 2D MOTs have been shown to lead to high atom flux, while ensuring a small footprint, in particular if atoms can be introduced into the system via dispensers instead of effusive ovens \cite{demarco1999an,moore2005collimated,jollenbeck2011hexapole}. However, for atomic species with low vapor pressure, such as Sr and Yb, sources that are solely based on a 2D MOT are not widely used, yet. Recent work has shown a Sr 2D MOT with a flux of up to $10^8$ atoms per second\cite{nosske2017two,barbiero2020sideband}, but the system requires an oven to be heated to about 500 $^\circ$C. Focusing on setups based on compact dispensers, a 2D MOT for Yb with a flux of $10^7$ atoms per second has been realized in a highly customized setup \cite{dorscher2013creation}. For Sr, dispenser-based 2D MOT designs have been demonstrated with a flux of $10^5$ atoms per second \cite{kock2013magneto, he2017towards}.

\begin{figure*}
    \centering
    \includegraphics[width=\textwidth]{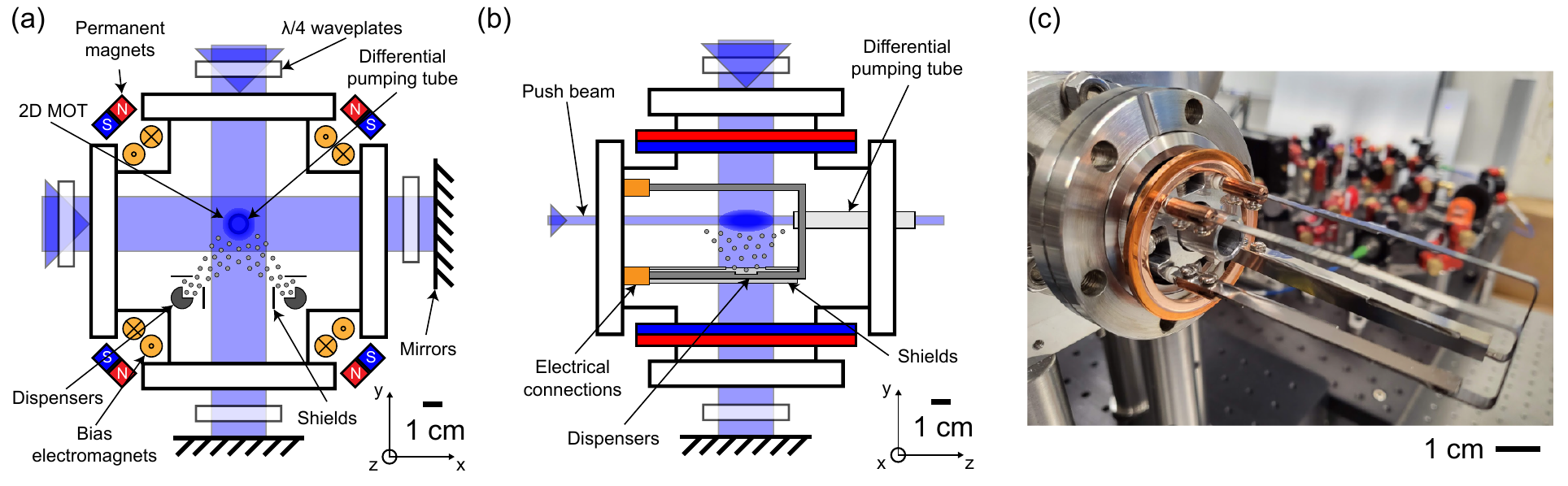}
    \caption{Overview of the technical implementation of the Sr 2D MOT. Schematic of the setup showing an axial view (a) and side view (b). (c) Image of the dispensers and the mounting structure. The cylinder in the middle is the end of a mounting tube that holds the dispenser assembly.}
    \label{fig:overall setup}
\end{figure*}

In this letter, we demonstrate a dispenser-based 2D MOT for Sr with a cold atom flux of up to $10^{8}$ atoms per second. The flux is measured via the loading rate of a 3D MOT, which constitutes a conservative lower bound for the cold atom flux produced by the 2D MOT. The setup is compact, maintenance free, consumes minimal electrical power due to the use of permanent magnets, and does not require a mechanical shutter to stop the atomic flux out of the source. While the performance is about an order of magnitude below Zeeman slowers, for most uses, such as optical lattice clocks and optical tweezer arrays (requiring 100 to 10,000 atoms), the performance is sufficient. Straightforward modifications, discussed at the end of the letter, should allow reaching a flux similar to Sr Zeeman slowers. Our setup is particularly suited for applications with low SWaP (size, weight, and power) requirements.

\section{Setup}

We start with a technical overview of the setup. Figure~\ref{fig:overall setup} shows the Sr 2D MOT design, illustrating the compact vacuum system and dispenser assembly.

\subsection{Vacuum system}
The vacuum system is mostly constructed from commercially available ultra-high vacuum (UHV) components. The 2D-MOT chamber is a six-way cross made of non-magnetic stainless steel (316SS). The vacuum is maintained by an ion pump with a pumping speed of 20 l/s. The 2D-MOT chamber is connected to the science chamber of the main apparatus through an exit port. The exit port is comprised of a tube that is 90 mm long with an inner diameter of 2 mm and serves as a differential pumping tube, allowing for a pressure differential of about $10^{4}$ between the 2D MOT and the science chamber. The bore is vertically offset 3 mm above the center of the six-way cross to account for the gravitational drop of the atomic beam on the way to the science chamber. Transverse cooling light enters from four uncoated Kodial glass viewports on the sides. The axial flange opposite to the exit port is designed to accept mounting structures for the dispensers and the electric connections, and has a through hole for the push beam. 

The design allows for a relatively short distance between the 2D MOT and the science chamber. The 2D MOT is formed about 1 cm away from the opening of the differential pumping tube and exits the differential pumping tube after about 10 cm of travel. The total travel distance between the 2D MOT and the center of the 3D MOT is 43 cm. This is substantially shorter than the 75 cm in an earlier realization of a Sr 2D MOT \cite{kock2013magneto}. Closer proximity increases the usable atomic flux as the atomic beam fans out less due to transverse motion on the way to the science chamber. This issue is more pronounced than in alkali 2D MOTs, as the transverse temperature of Sr remains relatively high (about 1 mK). In our setup, the solid angle of the atomic beam that can be captured is 126 mrad. We discuss below how this can be further increased.

During operation of the 2D MOT, the pressure in the 2D-MOT chamber remains as low as $1\times10^{-9}$ torr due to the low vapor pressure of Sr.

\subsection{Dispenser assembly}

Sr atoms are introduced into our system by generating a hot atomic jet that emerges from a resistively heated dispenser, containing bulk atomic Sr. The dispenser assembly is custom-designed. An important design criterion is to bring the output opening of the dispensers as close as possible to the 2D MOT trapping region  (see Fig.~\ref{fig:overall setup}). This allows for direct capture of atoms from the dispenser jet, minimizing the amount of atoms that fly-by uncaptured and stick to the chamber walls.

To this end, we utilize two U-shaped dispensers produced by a commercial vendor (AlfaVakuo) (see Fig.~\ref{fig:overall setup} (c)). They are comprised of a steel tube, filled with bulk Sr with natural abundance; the opening is a 5 mm long slit that before activation is sealed with indium. For the data reported here, we use dispensers with 2 mm diameter and a filling of 40 mg of Sr. Larger capacity dispensers with a filling of more than 200 mg of Sr can be accommodated with a similar design. The distance between the output opening and the 2D-MOT trapping region is about 1.5 cm. For electrical connection, the flat legs of the dispensers are connected to BeCu in-line barrel connectors that are isolated from the vacuum flange with ceramic spacers (FTACERB068, Kurt J.~Lesker).

In order to block the hot atom jet from coating the view ports, we have placed L-shaped shields around the dispensers made from stainless steel (SAE 304) sheet metal (see Fig.~\ref{fig:overall setup} (c)). The shields have a cut-out that restricts the solid angle of the fanned-out hot atom flux; the cut-out is narrow enough to protect the viewports from Sr coating and large enough to fully expose the trapping region. 

\begin{figure}
\centering
\includegraphics[width=.45\textwidth]{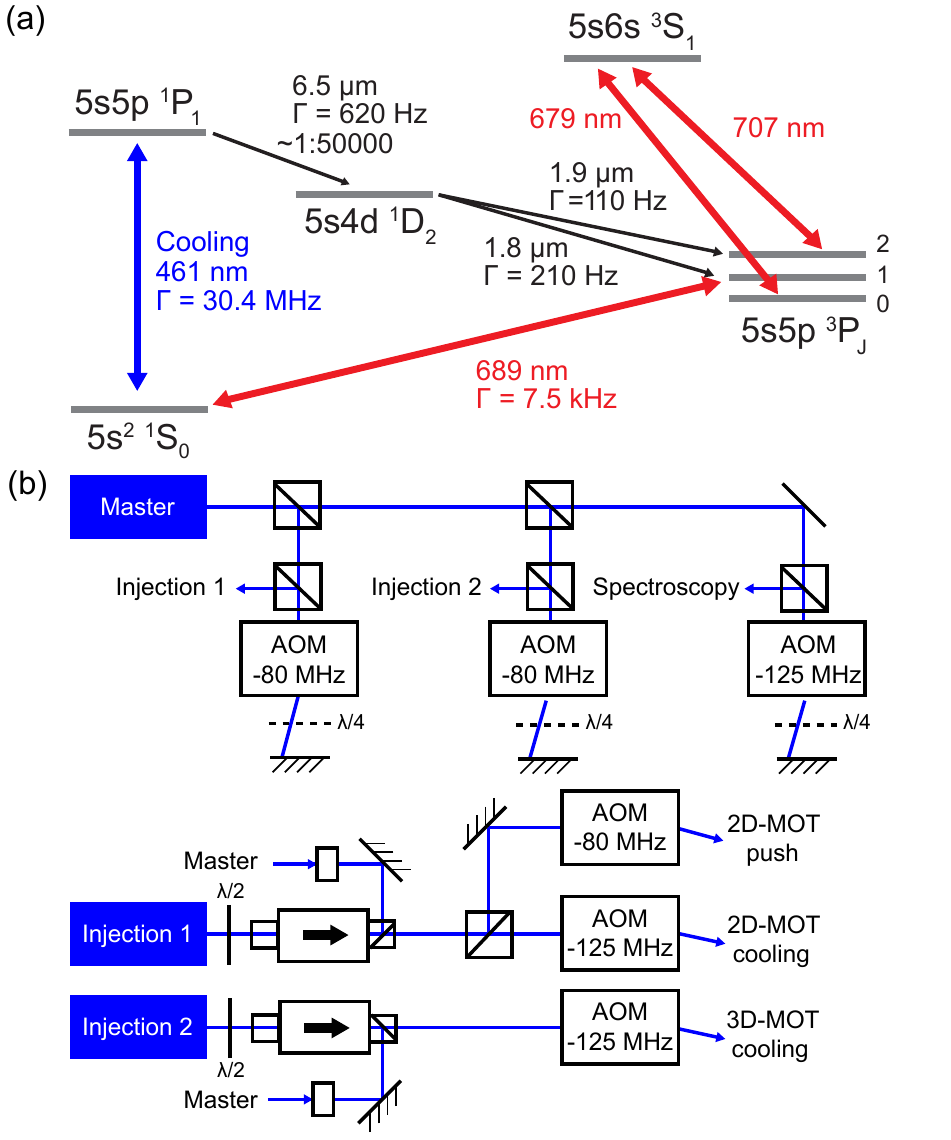}
\caption{Laser cooling of Sr. (a) Atomic levels and relevant optical transitions for cooling and repumping in our setup. (b) Schematic of the system for 461 nm laser light. A master laser, which is stabilized to a Sr spectroscopy cell, provides light for two injection-locked lasers. These lasers provide the optical power at 461 nm for the 2D and 3D MOT, respectively. The output of injection laser 1 is split into two beam paths for the 2D MOT and the push beam.}
\label{fig:SrCooling}
\end{figure}

\subsection{Laser System}

The relevant transitions\cite{boyd2007strontium, trautmann2022magnetic} for laser cooling and repumping of $^{88}$Sr are shown in Fig.~\ref{fig:SrCooling} (a). For the operation of the 2D MOT only laser light at 461 nm is used; for the operation of the 3D MOT we also use repumping light at 679 nm and 707 nm.

The 461 nm laser system consists of two diode lasers that are injection-locked to a master laser. The master laser is a commercial external cavity diode laser (ECDL) (Toptica DL Pro) stabilized to a hollow cathode lamp via polarization spectroscopy \cite{shimada2013simplified}. It injects two 500 mW diodes (Nichia NDB4916E), similar to the approach described in Ref.~\cite{schkolnik2020generating}. Each diode is housed in a temperature-stabilized mount (Thorlabs LDM56F) with a collimation lens (Thorlabs C330TMD-A, $\mathrm{f}=3.1$ mm, $\mathrm{NA}=0.7$). Injection happens via an optical isolator (Newport ISO-04-461-MP). With a few mW of seed power, the lasers stay stably locked. The repumping transitions at 679 nm and 707 nm are addressed with laser light from ECDLs that are stabilized to a high-precision wavemeter (HighFinesse WS-7).

The 2D MOT is operated with a total power of 150 mW of 461 nm light, equally distributed onto the two retro-reflected arms (see Fig.~\ref{fig:overall setup} (a)). The light is delivered to the setup using a polarization maintaining fiber (OZ Optics QPMJ-3A3A-400). The beams are shaped to a $1/e^2$-radius of 6 mm using an outcoupler lens ($\mathrm{f}=8$ mm) and a magnifying telescope ($\mathrm{f}=25$ mm and $\mathrm{f}=200$ mm). The push beam is typically operated at a power of $50 - 100\, \mu$W and has a $1/e^2$-radius of 0.8 mm.

\subsection{Magnetic field generation}

The quadrupole magnetic field for the 2D MOT is generated via permanent magnets, providing the necessary field gradients without consuming power. They are screwed onto a slender aluminum mount that is attached to a robust 3D-printed mount that allows for position adjustments of the magnets. We use four rectangular permanent magnets (N45 3"x1/2"x1/4", CMS Magnetics). Mechanical tuning of the magnet location allows adjustments of the field gradient between 20 and 200 G/cm at the trapping region. The gradients are measured prior to installation of the magnet assembly and match the simulated field distribution. We find optimal performance for a magnetic field gradient of 64 G/cm, but the 3D MOT loading rate remains relatively insensitive over a broad range, i.e., 50-150 G/cm. To allow for fine positioning of the 2D MOT location with respect to the differential pumping tube, additional Helmholtz coil pairs are wound around the ports of the six-way cross using standard magnet wire. 

\begin{figure}
\centering
\includegraphics[width=.45\textwidth]{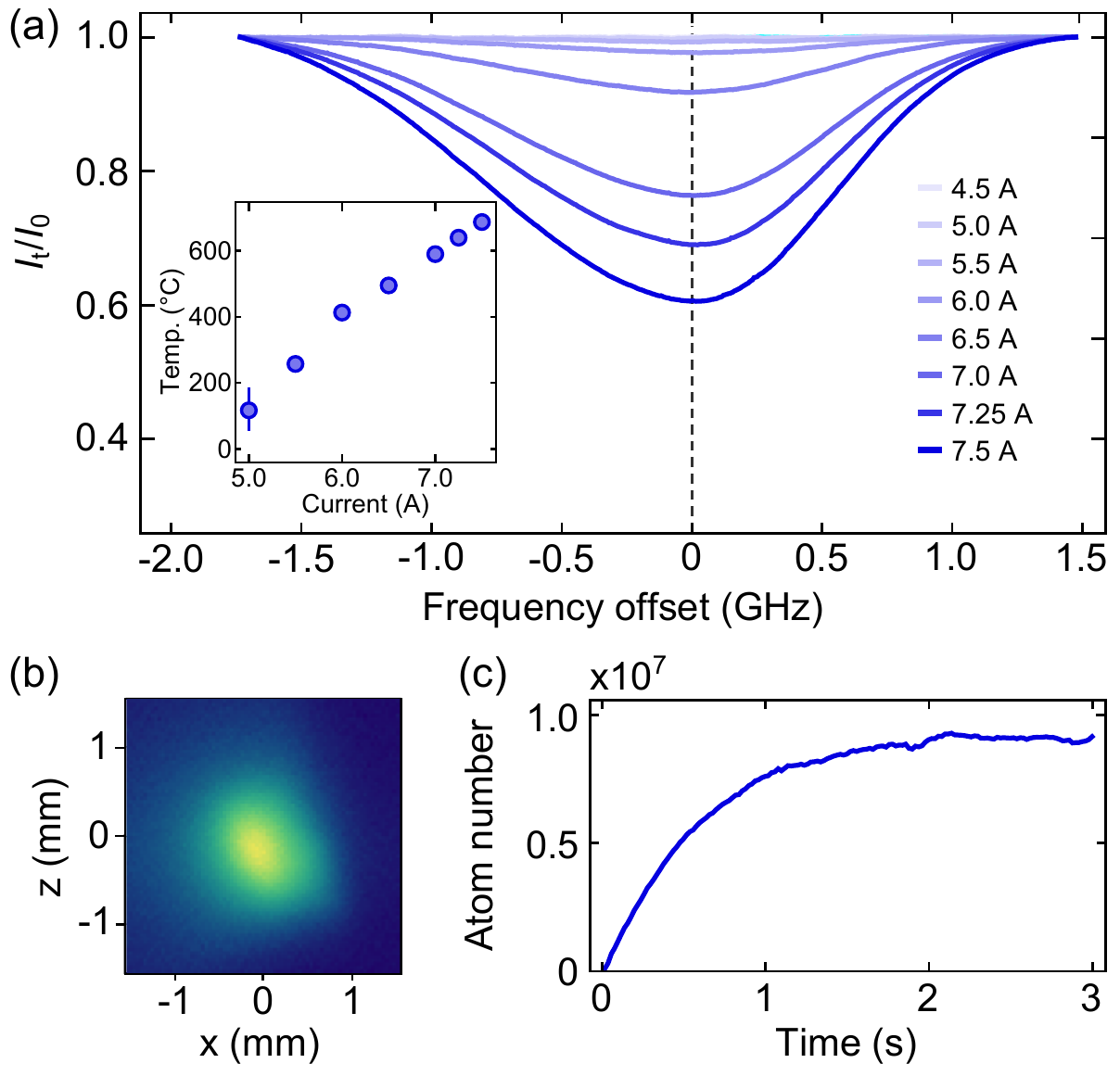}
\caption{Calibration of 2D and 3D MOT parameters. (a) Doppler spectroscopy on the atom jet. Inset shows the temperature of the dispensers as a function of dispenser current. The asymmetry of the Doppler profiles likely arises from absorption on the 461 nm transitions of the isotopes $^{86}$Sr and $^{87}$Sr that are slightly red-detuned compared to the dominant isotope  $^{88}$Sr. (b) Fluorescence image of the 3D MOT after loading, corresponding to about $10^7$ Sr atoms. (c) Loading curve of the 3D MOT.}
\label{fig:3}
\end{figure}

\section{Characterization}

First, we characterize the temperature of the hot atom jet out of the Sr dispensers as a function of dispenser current. To this end, we perform Doppler spectroscopy on the atom jet using the on-axis push beam. We assume that in the direction of the push beam the velocity distribution of the atoms is well-approximated by a one-dimensional Maxwell-Boltzmann distribution. By fitting the measured Doppler profiles, we obtain an approximate value for the dispenser temperature at different currents (see Fig.~\ref{fig:3} (a)).

In the following, we characterize the performance of the 2D MOT by measuring the loading rate of the 3D MOT. The loading rate of the 3D MOT is a measure of the trappable flux and smaller than the total atom flux out of the 2D MOT. As such, it is a conservative lower bound for the cold atom flux from the source. The 3D MOT is comprised of three retro-reflected beam pairs and a magnetic quadrupole coil with its symmetry axis aligned vertically. The horizontal (vertical) beams use a power of 4 mW (2.5 mW) and have a  $1/e^2$-radius of 3.5 mm (2.5 mm). The detuning of the cooling beams is $-1.5 \,\Gamma$. The magnetic quadrupole field has a gradient of 45 G/cm along the vertical axis. The atom number in the 3D MOT is evaluated via fluorescence imaging during loading using an \mbox{EMCCD} camera (Andor iXon Ultra 888) (see Fig.~\ref{fig:3} (b)). The atom number calibration of fluorescence imaging has been confirmed via absorption imaging. We extract the loading rate $L$ by fitting the observed MOT loading curves (an example is shown in Fig.~\ref{fig:3} (c)) to the solution of the differential equation $\dot{N}(t) = - \alpha N (t) + L$, where \(N(t)\) is the atom number at time $t$ and \(\alpha\) is the single-body loss rate.

\begin{figure}
\centering
\includegraphics[width=.45\textwidth]{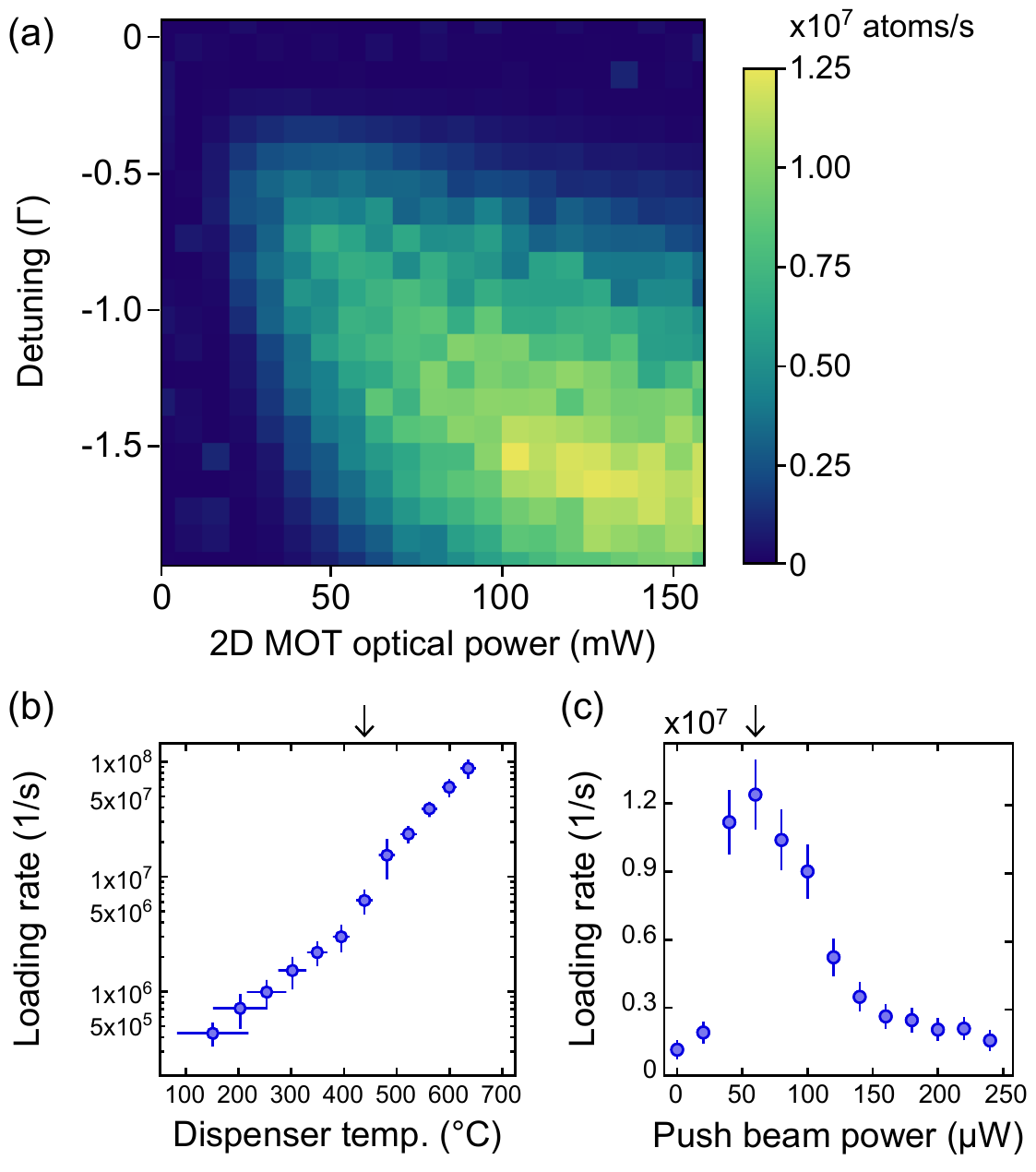}
\caption{Performance of the 2D MOT system. (a) Optimization of the achievable loading rate as a function of laser detuning and laser power. The dispenser temperature is 440 $^\circ$C (b) Loading rate as a function of dispenser temperature. The arrow marks the temperature used for panel (a). (c) Loading rate as a function of push beam power. Arrow marks the push beam power used for panel (a). Data sets in (b) and (c) are recorded at a detuning of $-1.5\,\Gamma$ and laser power of $150(2)$ mW.}
\label{fig:4}
\end{figure}

We study the performance of the 2D MOT as a function of the relevant experimental parameters. First, we scan the detuning and optical power of the cooling light, as shown in Fig.~\ref{fig:4} (a). We find peak performance for a detuning of $-1.5\,\Gamma$ and an optical power of $120-150$ mW. Near-peak performance is observed in a fairly broad parameter range of detuning and optical power. In Fig.~\ref{fig:4} (b), we show that the atomic flux can be smoothly tuned over several orders magnitude via the dispenser temperature. For the highest dispenser temperature at 635 $^\circ$C we obtain a loading rate of up to $10^{8}$ s$^{-1}$. In daily operation, we use a temperature of 440 $^\circ$C, which provides us with a loading rate and 3D MOT size that is sufficient for our Sr tweezer experiment. Finally, we investigate the loading rate as a function of push beam power as shown in Fig.~\ref{fig:4} (c). We observe a pronounced peak at around 60 $\mu$W, which corresponds to an intensity of $I= 0.14 \,I_{\mathrm{sat}}$. For lower powers, the atomic beam is not effectively pushed through the differential pumping tube and fans out too much before reaching the 3D MOT region. For higher powers, the atomic beam is accelerated to velocities beyond the capture velocity of the 3D MOT at about $30$ m/s. 

The atomic flux out of the 2D MOT can be effectively stopped by simultaneously switching off the cooling and push beams. An additional mechanical shutter is not needed.

\section{Discussion}
Compared to dispenser-based 2D MOT Sr sources reported previously \cite{kock2013magneto, he2017towards}, our system shows an enhancement of cold atom flux by three orders of magnitude. We attribute this improvement to the close proximity of the dispensers to the 2D MOT cooling region, facilitating efficient capture from the dispenser jet and minimizing the amount of atoms that fly by uncaptured. Compared to Sr sources based on an oven \cite{nosske2017two}, our system achieves a slightly lower flux, but significantly reduces the size and complexity by replacing the oven with dispensers and eliminating the Zeeman slower. Due to the small heat capacity of the dispenser assembly, the atomic flux out of the dispenser can also be switched on and off on the second-scale by controlling the current, compared to tens of minutes for an oven. The total electrical power consumption of the setup is 13 W (including dispensers and shim coils, without lasers), which is ideal for the use in setups with stringent SWaP requirements. The presented source can also be useful for applications in which blackbody radiation needs to be suppressed, e.g., for precision measurements on the Sr clock transition \cite{nicholson2015systematic} and when using Rydberg states \cite{gallagher1988rydberg}, as the hot dispensers do not have a direct line-of-sight with the science chamber. We note that, in addition to $^{88}$Sr, we have also observed cooling and trapping of all other naturally occurring isotopes of Sr with the observed flux scaled by the respective percentages of the natural abundance.

Straightforward modifications should allow further improvement of performance. Fig.~\ref{fig:4} (a) suggests that a higher optical power can further enhance the achievable flux. As demonstrated in Ref.~\cite{barbiero2020sideband}, the addition of a sideband to the cooling laser at a detuning of $-3 \Gamma$ promises to further enhance the flux by up to a factor of four. The dispensers, which currently hold 40 mg of atomic Sr, can also be replaced with 200 mg dispensers, which will allow a five times higher flux at identical lifetime of the dispensers \footnote{We have used the 40 mg dispensers for ten months in daily operation and not seen a degradation of performance.}. Finally, 3D MOT loading can be further enhanced by shortening the distance between the 2D MOT and the 3D MOT, which currently is about 40 cm. For example, the glass cell science chamber in our setup has an approximately 10 cm-long glass metal transition, which can be removed in a modified setup. This change by itself should allow reaching a loading rate of $10^8$ s$^{-1}$ at dispenser temperatures below 500 $^\circ$C.

\section{Conclusion}
We have demonstrated a cold atomic beam source for Sr based on a dispenser-loaded 2D MOT. The setup loads a 3D MOT with a rate of up to $10^{8}$ atoms per second, while being compact, robust, and power efficient in operation. With the described modifications, the setup has the potential to reach a loading rate of $10^{9}$ atoms per second and beyond.  It may find uses in Sr-based quantum simulation, quantum computing, and optical clock devices, in particular for field- or space-deployable designs with stringent size and power constraints. Going beyond Sr, we expect that a similar design approach can be employed to realize 2D MOT sources for other alkaline-earth and lanthanide atoms.

\section{Acknowledgements}
We thank Weijun Yuan and Siwei Zhang for experimental assistance. This work was supported by an NSF QII-TAQS Award (Award No. 1936359) and an NSF Convergence Accelerator Award (Award No. 2040702). S.W.~acknowledges additional support from the Alfred P. Sloan Foundation.


%

\end{document}